\documentclass{desyproc}

\usepackage{dsfont}
\usepackage{epstopdf}

\begin{document}
\title{
Distribution Amplitudes of Heavy Hadrons: \\ 
Theory and Applications
}

\author{{\slshape Alexander Parkhomenko}\\[1ex]
P.\,G.~Demidov Yaroslavl State University, Yaroslavl, Russia}

\contribID{xy}

\desyproc{DESY-PROC-2016-04}
\acronym{HQ2016} 

\maketitle

\begin{abstract}
The physics of heavy quarks becomes a very reach area 
of study thanks to an excellent operation of hadron 
colliders and $B$-factories and exciting results from them. 
Experimental data obtained allows to get some information 
about the heavy hadron dynamics. In this case, the models 
for the heavy hadron wave-functions are required to do 
theoretical predictions for concrete processes under study. 
In many cases, the light-cone description is enough 
to obtain theoretical estimates for heavy hadron decays. 
A discussion of the wave-functions of the $B$-meson and 
heavy bottom baryons in terms of the light-cone distribution 
amplitudes is given in this paper. Simple models for the 
distribution amplitudes are presented and their scale 
dependence is discussed. Moments of the distribution 
amplitudes which are entering the branching fractions 
of radiative, leptonic and semileptonic $B$-mesons decays 
are also briefly discussed.   
\end{abstract}

\section{Introduction}
\label{sec:Parkh-Introduction}

Physics of heavy hadrons, both experimental and theoretical, 
still remains a hot topic at present. From one side, the  
remarkable operation of the hadron collider LHC at CERN provides 
us with a lot of interesting and exciting results and, from 
the other side, many more results can be obtained from the 
$B$-factory at KEK which is under construction now and hopefully 
to be run in a year from now. Among the excellent achievements 
obtained at the LHC, there is the measurement of the rare 
purely leptonic decay $B_s \to \mu^+ \mu^-$ by the LHCb and 
CMS collaborations~\cite{CMS:2014xfa} in a nice agreement 
with the theoretical expectations based on the Standard Model 
(SM) (see, for example,~\cite{Bobeth:2013uxa} and reference 
therein) and the evidence of the similar decay $B^0 \to \mu^+ \mu^-$ 
seen by the same collaborations~\cite{CMS:2014xfa} 
is also in agreement with the SM predictions~\cite{Bobeth:2013uxa}.  
Let us note that both decays were considered as a clean candle 
into possible new physics at the flavor-physics frontue, when predicted 
branching fractions are substantially larger the SM expectations.   
At the statistics have been already collected at the LHC, it 
starts possible to make a detail combine analysis of the rare 
$B \to K^{(*)} \ell^+ \ell^-$ decays, where $\ell = e, \, \mu$,
in the lepton-pair invariant mass squared and spherical angles. 
In particular, some quantities show a sizable deviation from 
theoretical predictions and these require theoretical explanation. 
The other semileptonic decay $B^+ \to \pi^+ \mu^+ \mu^-$ have been also 
observed at the LHC~\cite{LHCb:2012de} and, on the Run-II statistics, 
for the first time partially integrated branching fraction over 
the lepton-pair invariant mass divided into 8~bins have been 
experimentally obtained~\cite{Aaij:2015nea} in a good agreement 
with the SM predictions. All these results need to be approved 
at the $B$-factory SuperKEKB  
after its operation will start and, in addition, one should wait  
new exciting results as several $B$-meson decay modes, being 
unobservable at the LHC, can be measured by the Belle collaboration 
and are of great importance to get a complete picture of $B$-meson 
physics. Among them, one can specify the decays like 
$B \to K^{(*)} \nu \bar \nu$ and $B^0 \to \pi^0 \ell^+ \ell^-$. 

In a difference to $B$-meson decays, the situation with bottom 
baryons is a little bit worser as the SuperKEKB machine is not 
designed for the $\Lambda_b$ and heavier baryon production. 
So, the LHC is the only source for the bottom-baryon study for quite 
some time. As the result, only limited information about this 
part of the hadronic sector is available and a lot of theoretical 
predictions will wait their check for future experimental facilities. 
Interesting processes are the rare semileptonic decays  
$\Lambda_b \to \Lambda \ell^+ \ell^-$ decays, where $\ell = e, \, \mu$, 
which are the baryonic realization of the flavor-changing neutral 
current transition $b \to s \ell^+ \ell^-$, quite sensitive 
to induced by loop diagrams contributions from new physics.  

Theoretical predictions for weak decays require some information 
about the dynamics of light quarks in heavy hadrons~--- 
$B_{(s)}$-meson and bottom baryon. Several approaches have been 
already worked out and, in some cases, experimentally checked 
like $B \to K^{(*)} \ell^+ \ell^-$ decays, where $\ell = e, \, \mu$, 
both at $B$-factories at SLAC and KEK and at the hadron colliders 
at FNAL and CERN. Among them, factorization approaches are 
of priority. In particular, the Soft-Collinear Effective Theory 
(SCET) originally suggested for making prediction based on the idea 
of the energy separation of light degrees of freedom in weak decays,  
have become a powerful tool for other multiscale processes which 
are under intensive study at the LHC now. As a byproduct of this theory, 
one needs to know several moments of the heavy-hadron wave-function 
which are entering invariant amplitudes of decays. Moreover, even 
a shape of the heavy-hadron wave-function is required to determine 
the momentum-squared dependence of the transition form factors. 
So, a business connected with a study of the dynamical properties 
of the heavy-hadron wave-functions is a useful direction in 
theoretical high-energy physics and its basis and some applications, 
this lecture is devoted.

\section{Light-Cone Distribution Amplitudes of $B$-Meson}
\label{sec:Parkh-LCDAs-B-meson}

\begin{figure}[tb] 
\centerline{
\includegraphics[width=0.75\textwidth]{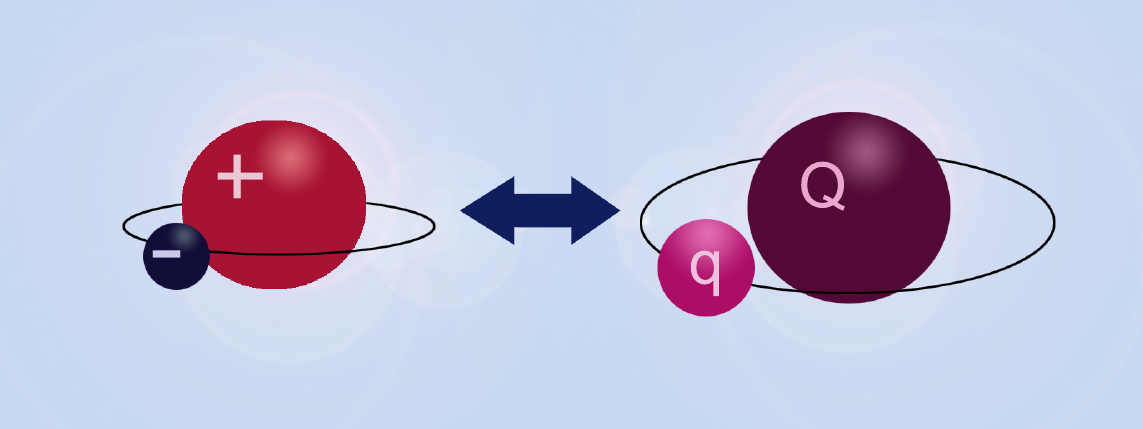} 
} 
\caption{
The similarity between the hydrogen atom (left) 
and heavy meson (right). 
}
\label{fig:HA-HM-similarity}
\end{figure}

The dynamics of a light quark inside a heavy meson is 
convenient to describe within the Heavy Quark Effective Theory 
(HQET)~\cite{Manohar:2000dt,Grozin:2004yc,Mannel:2004ce}.  
In this approach, the heavy antiquark is considered 
as an external source (either static or slowly moving 
in dependence on the frame specified) and the light quark 
completely determines the meson properties. So, HQET 
is a useful tool in theoretical analysis of heavy hadrons.  
As the heavy meson is very similar to the hydrogen atom 
(see Fig.~\ref{fig:HA-HM-similarity}), one should use 
the effective mass of heavy meson~$\bar\Lambda = m_M - m_Q$ 
which is nothing else but the difference between 
the meson mass~$m_M$ and the heavy-quark mass~$m_Q$.   
Applying the pole scheme for the heavy antiquark, 
one can get the estimate $\bar\Lambda \simeq 0.5$~GeV 
for the lowest-mass $D$- and $B$-mesons~\cite{Olive:2016xmw}.  
The analogy between the heavy meson and hydrogen atom 
can be extended even further taking into account the fact 
that in the limit of an infinitely large mass of the 
heavy antiquark the heavy-quark spin does not influence 
the light-quark dynamics and, so, decouples. Such 
an approximation is known as the Heavy-Quark-Symmetry 
(HQS) limit in which the heavy quark can be considered 
as a spinless particle. In working out the dynamical 
properties of the meson, one can assume the heavy quark 
to be a scalar particle and study the so-called ``spinor'' 
meson~\cite{Grozin:1996pq}.  
The realistic quantum numbers of the meson and its 
wave-function can be obtained after a contraction 
with the heavy-quark spin.

The bilocal operator interpolating the heavy meson in the 
valence quark-antiquark approximation (the lowest Fock state) 
in the HQS limit is as follows~\cite{Grozin:1996pq}: 
%
\begin{equation}
\widetilde O (z) = Q^* (0) \, {\cal P} \exp \left \{ - i g_{\rm st} 
\int_0^z dz'^\mu A^a_\mu (z') \, \frac{\lambda^a}{2} \right \} q (z) 
= Q^* (0) \, E (0,z) \, q (z) . 
\label{eq:BO-def}
\end{equation}
The link between the quarks is the path-ordered exponential 
called the Wilson line~\cite{Peskin:1995ev},   
%
%
where $A^a_\mu (z)$ is the gluonic field,  
$g_{\rm st}$ is the strong coupling, and  
$\lambda^a$ ($a = 1, \ldots, 8$) are the 
Gell-Mann matrices. Sometimes it is convenient 
to use the notation: $A_\mu = A^a_\mu \lambda^a/2$.  
%
%
%

In many exclusive decays of the $B$-meson only one projection 
of the light-quark momentum gives the dominant contribution 
into the decay amplitude. So, it is necessary to find such 
a direction and study the dependence of the meson quantities 
on the corresponding momentum projection. On this way, the 
light-cone representation of four-vectors appears to be very 
convenient. To get a corresponding decomposition, one needs to 
introduce a light cone and specify two light-like four-vectors 
on it, for example,   
$n_\pm^\mu = \left ( 1, 0, 0, \mp 1 \right )/\sqrt 2$, 
where $n_\pm^2 = 0$ and $(n_+ n_-) = 1$. 
They can be related with physical vectors presented 
in a problem. In particular, if one assumes that the 
light quark in the bilocal operator~(\ref{eq:BO-def}) 
is a massless particle situated on the light cone 
and the gauge link is a straight line connecting its 
position with the origin, the four-vector~$n_+^\mu$ 
can be directed along this link. In this case, 
$dz'^\mu = dz'_- n_+^\mu$ in~(\ref{eq:BO-def}) 
and the variable~$z'_-$ specifies the position 
of the gluon field on the line. The decomposition 
of an arbitrary four-vector is as follows: 
%
\begin{equation}
A^\mu = A_+ n^\mu_- + A_- n^\mu_+ + A_\perp^\mu ,  
\label{eq:4-vector-decomposition}
\end{equation}
where $A_\pm = n^\mu_\pm A_\mu$ are two light-cone projections.  
%
%
Note that the scalar product $(A^a (z') dz') = A_+^a (z'_-) dz'_-$ 
in the exponential in~(\ref{eq:BO-def}) and turns out to be zero  
in the Fock-Schwinger gauge $A_+^a (z'_-) = 0$. In this gauge, 
adopted further in this paper, the gauge link becomes trivial, 
$E (0,z) = 1$, and the expression for the meson bilocal 
current~(\ref{eq:BO-def}) simplifies.


The meson-to-vacuum transition matrix element 
at $z^2 = 0$~\cite{Grozin:1996pq} is of our interest: 
%
\begin{equation} 
\langle 0 \vert Q^* (0) \, q (z) \vert M (v) \rangle = 
f_M \left\{ \tilde\varphi_+ (t) + 
\left [ \tilde\varphi_- (t) - \tilde\varphi_+ (t) \right ] 
\frac{\hat z}{2 t} \right\} U (v) , 
\label{eq:ME-spinor-like}
\end{equation}
where $f_M$ is a constant with the dimension of a mass,  
$t = (v z)$ is the time in the heavy-meson rest frame,     
where the meson four-velocity is $v^\mu = (1, 0, 0, 0)$,
the bispinor $U (v)$ is the non-relativistic wave-function 
of the ``spinor-like'' meson, and $\tilde\varphi_\pm (t)$ 
are the distribution amplitudes (DAs) of the heavy meson.  
The Fourier transforms of DAs are usually required 
in constructing matrix elements of heavy meson decays:  
\begin{equation} 
\tilde\varphi_\pm (t) = 
\int_0^\infty d\omega \, {\rm e}^{-i \omega t} \phi_\pm (\omega) . 
\label{eq:B-DAs-Fourier-transforms}
\end{equation}

In performing calculations, it is convenient to introduce 
the projection operator onto the $B^{(*)}$-meson state.  
Note that the heavy $B$- and $B^*$-mesons are degenerate 
in this approach.  
The $B$-meson-to-vacuum transition matrix element 
at $z^2 = 0$ in the real world after the heavy-quark 
spin is taken into account can be written in the 
form~\cite{Grozin:1996pq,Beneke:2000wa}:
%
\begin{eqnarray} 
\langle 0 \vert 
\bar q_\alpha (z) \, h_{v, \beta} (0) 
\vert \bar B^{(*)} (v) \rangle = 
\eta_{B^{(*)}} \frac{i f_{B^{(*)}} m_{B^{(*)}}}{4} 
\left [ \left ( 1 + \hat v \right ) \left \{ \tilde\varphi^B_+ (t) - 
\left [ \tilde\varphi^B_+ (t) - \tilde\varphi^B_- (t) \right ] 
\frac{\hat z}{2 t} \right \} \gamma_5 (\hat \varepsilon) 
\right ]_{\beta\alpha} \!\!\!,  
\label{eq:B-PO-two-particle}   
\end{eqnarray} 
where $\varepsilon^\mu$ is the polarization vector 
of the $B^*$-meson, $\eta_B = -1$ for the $B$-meson 
and $\eta_{B^*} = 1$ for the $B^*$-meson. 
The HQS gives the relations $f_{B^*} = f_B$ and 
$\tilde\varphi^{B^*}_\pm (t) = \tilde\varphi^B_\pm (t)$. 

The projection operator onto the three-particle state 
can be also determined through the $B$-meson-to-vacuum 
transition matrix element at $z^2 = 0$ both for the 
``spinor-like'' meson and after the spin of the heavy 
quark is switched on. For this matrix element one needs to 
introduce four distribution amplitudes~\cite{Kawamura:2001jm}:  
$\tilde\Psi_A^B (t, u)$, $\tilde\Psi_V^B (t, u)$, 
$\tilde X_A^B (t, u)$, and $\tilde Y_A^B (t, u)$.   
The corresponding projection operator 
$\langle 0 \vert \bar q_\alpha (z) \, G_{\lambda\rho} (u z) \, 
h_{v, \beta} (0) \vert \bar B (v) \rangle$ was also worked out 
and can be found in~\cite{Khodjamirian:2006st}:  
%
A similar projection operator for $B^*$-meson-to-vacuum 
transition matrix element at $z^2 = 0$ can be easily 
obtained from the above one after making the replacement 
$\gamma_5 \to \hat\varepsilon^*$ as in~(\ref{eq:B-PO-two-particle}).   
%

Equations of motion (EoM) for heavy and light quarks 
which are assumed to be on the mass shell result into 
relations among the $B$-meson DAs~\cite{Kawamura:2001jm}:  
%
\begin{equation}
\phi^B_+ (\omega) + \omega \, \frac{d\phi^B_- (\omega)}{d\omega} = I (\omega) , 
\quad 
\left ( \omega - 2 \bar\Lambda \right ) \phi^B_+ (\omega) + 
\omega \, \phi^B_- (\omega) = J (\omega) ,  
\label{eq:DAs-relations}
\end{equation}
where $I (\omega)$ and $J (\omega)$ are determined 
by the three-particle quark-antiquark-gluon DAs~\cite{Kawamura:2001jm}.  
The Wandzura-Wilczek~\cite{Grozin:1996pq} relation follows from 
the first equation in~(\ref{eq:DAs-relations}) when $I (\omega) = 0$: 
\begin{equation}
\phi^B_- (\omega) = \int_{\,\omega}^\infty 
\frac{\phi^B_+ (\omega')}{\omega'} \, d\omega' . 
\label{eq:Wandzura-Wilczek-limit}
\end{equation}

The basic property of the $B$-meson DAs is their scale dependence  
and one needs to know corresponding evolution equations. 
For the leading $B$-meson DA, this equation was worked out 
by B.~Lange and M.~Neubert~\cite{Lange:2003ff}:  
%
\begin{equation}
\frac{d \phi^B_+ (\omega; \mu)}{d\ln\mu} = 
- \frac{\alpha_{\rm st} (\mu) \, C_F}{\pi} \int_0^\infty d\omega' \, 
\gamma^{\rm LN} (\omega, \omega'; \mu) \, \phi^B_+ (\omega'; \mu) , 
\label{eq:phi-plus-evol-eqn}
\end{equation}
where $C_F = 4/3$. The Lange-Neubert anomalous dimension 
is as follows~\cite{Lange:2003ff}: 
\begin{eqnarray}
&& 
\gamma^{\rm LN} (\omega, \omega'; \mu) = 
\left ( \ln \frac{\mu}{\omega} - \frac{5}{4} \right ) 
\delta \left ( \omega - \omega' \right ) - 
\Gamma_{\rm LN} (\omega', \omega) , 
\label{eq:LN-kernel-gen} \\
%
&& 
\Gamma_{\rm LN} (\omega', \omega)  = 
\left [ \frac{\omega}{\omega'} \, 
\frac{\Theta \left ( \omega' - \omega \right )}{\omega' - \omega} + 
\frac{\Theta \left ( \omega - \omega' \right )}{\omega - \omega'}  
\right ]_\oplus .  
\label{eq:LN-kernel}
\end{eqnarray}
Here, the $\oplus$-convention is introduced: 
\begin{eqnarray}
\int_0^\infty  d\omega' f (\omega') 
\left [ \gamma (\omega', \omega) \right ]_\oplus = 
\int_0^\infty  d\omega' \left [ f (\omega') - f (\omega) \right ] 
\gamma (\omega', \omega) .  
\label{eq:O-plus-def}
\end{eqnarray}
It was also shown in~\cite{Lange:2003ff} that  
the Lange-Neubert kernel factorizes in the space of moments:  
\begin{eqnarray}
\tilde\Gamma_{\rm LN} (N) =  
\int_0^\infty  d\omega 
\left ( \frac{\omega}{\omega'} \right )^{N - 1}  
\Gamma_{\rm LN} (\omega', \omega)  = 
- \Psi (N) - \Psi (-N) - 2 \gamma_{\rm E} , 
\label{eq:LN-kernel-fact}
\end{eqnarray}
where $\gamma_{\rm E} = 0.577216$ is the Euler's constant and  
$\Psi (x) = \Gamma' (x)/\Gamma (x)$ is the digamma function. 
The analytic solution of the evolution equation can be written 
in the integral form as follows:  
\begin{equation}
\phi^B_+ (\omega; \mu) = \frac{1}{2\pi} \int_{-\infty}^{+\infty}
dt \, \varphi_0 (t) \, f (\omega, \mu, \mu_0, i t) .
\label{eq:phi-B-plus-solution}
\end{equation}
It depends on the function $f (\omega, \mu, \mu_0, i t)$ which 
can be calculated in the perturbation theory.  
The other function $\varphi_0 (t)$ is arbitrary and fixed 
by the condition $\varphi_0 (0) = \lambda_B^{-1}$ only. 
The shape of $\varphi_0 (t)$ can be determined after 
a model for $\phi^B_+ (\omega; \mu)$ is specified. 
%

%

\section{Applications of $B$-meson DAs}
\label{sec:Parkh-$B$-meson-applications}

The $B$-meson distribution amplitudes, 
usually in the form of inverse moments, are entering 
the exclusive decay amplitudes of $B$-mesons. 
In this section some processes where the distribution 
amplitudes are important, are presented and briefly discussed.

\subsection{$B^+ \to \ell^+ \nu_\ell \gamma$ Decay}  
\label{ssec:Parkh-B-to-ell-nu-gamma-decay} 

\begin{figure}[tb]
\begin{center}
\includegraphics[width=0.25\textwidth]{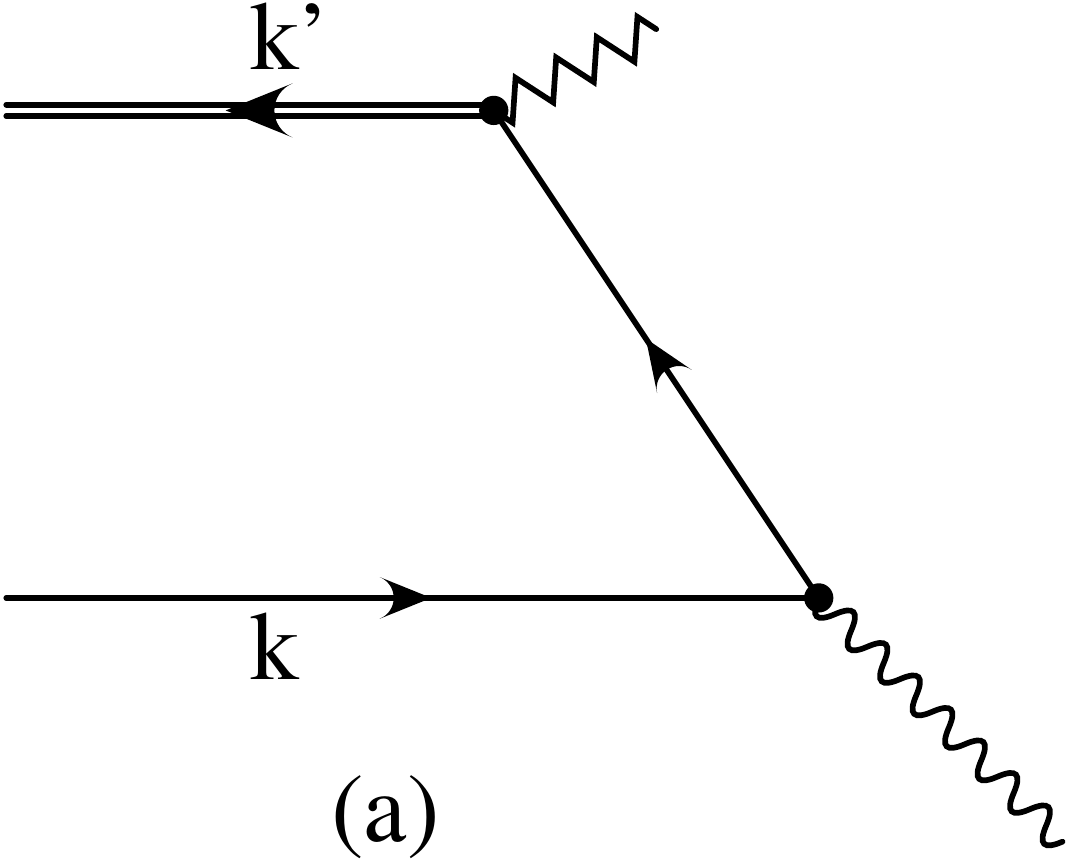} 
\qquad 
\includegraphics[width=0.25\textwidth]{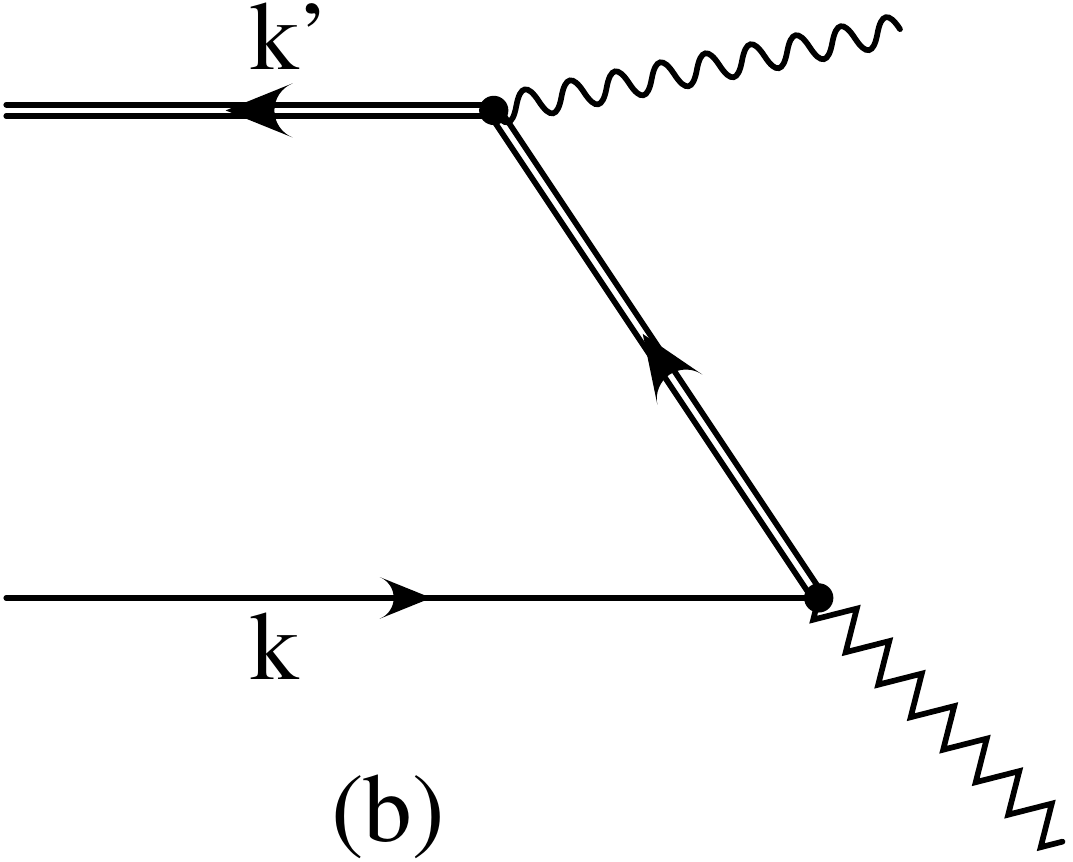} 
\end{center}
\caption{
The $B^+ \to \ell^+ \nu_\ell \gamma$ decay amplitudes 
in the leading order in perturbation theory.
} 
\label{fig:B-to-ell-nu-gamma-LO}
\end{figure}

Let us start the discussion with one of the most clean 
$B$-meson decay modes~--- the radiative leptonic 
$B^+ \to \ell^+ \nu_\ell \gamma$ decay. 
In the leading order in the perturbation theory, the decay 
amplitude is presented by two Feynman diagrams shown  
in Fig.~\ref{fig:B-to-ell-nu-gamma-LO} and can be written as follows: 
%
\begin{equation}
{\cal M} = \frac{4 G_F}{\sqrt 2} \, V_{ub}^* 
\left\langle \gamma (p, \epsilon) \left | \bar b \gamma_\mu 
\left ( 1 - \gamma_5 \right ) u \right | B^+ (v) \right\rangle 
\left [  
\bar u (q_\ell) \gamma^\mu \left ( 1 - \gamma_5 \right ) u (q_\nu) 
\right ] .
\label{eq:B-to-ell-nu-gamma-LO}
\end{equation}
%
%
%
The $B^+ \to \gamma$ transition matrix element entering 
this amplitude can be parameterized by two form factors 
$F_V (E_\gamma)$ and $F_A (E_\gamma)$~\cite{Korchemsky:1999qb}: 
%
\begin{eqnarray}
&& 
\left\langle \gamma (p, \epsilon) \left | \bar b (0) \gamma^\mu 
\left ( 1 - \gamma_5 \right ) u (0) \right | B^+ (v) \right\rangle 
\label{eq:B-to-gamma-TME} \\ 
&& 
= \sqrt{4\pi\alpha} \left \{ 
\varepsilon^{\mu \rho \sigma \tau} v_\rho p_\sigma \epsilon_\tau^* \, 
F_V (E_\gamma) - i \left [ 
(v p) \epsilon^{*\mu} - (v \epsilon^*) p^\mu 
\right ] F_A (E_\gamma) \right \} , 
\nonumber 
\end{eqnarray}
where $E_\gamma = (v p) = m_B \left ( 1 - q^2/m_B^2 \right )/2$  
is the photon energy. 
The kinematical region of the process is determined 
by the condition $E_\gamma \gg \Lambda_{\rm QCD}$, 
with $\Lambda_{\rm QCD}$ being the QCD parameter,
where perturbative QCD methods for exclusive processes 
can be applied.  

%
It was demonstrated that in this order the factorization 
approach is applicable and the most convenient tool for 
the QCD factorization implementation is the Soft-Collinear 
Effective Theory (SCET)~\cite{Bauer:2000yr,Bauer:2001yt,Beneke:2002ph}. 
%
We are not going to discuss this theory in details but 
give the following references~\cite{Becher:2014oda,Grozin:2016uyg}, 
where a detail discussion of this theory can be found.

The differential width of the $B^+ \to \ell^+ \nu_\ell \gamma$ 
decay calculated within the SCET can be written in the 
form~\cite{DescotesGenon:2002mw}:  
%
\begin{equation}
\frac{d\Gamma}{dE_\gamma} = 
\frac{\alpha G_F^2 f_B^2 |V_{ub}|^2 m_B^4}{54 \pi^2} \, 
\left ( C_3^{\rm SCET} \right )^2 
\frac{x_\gamma \left ( 1 - x_\gamma \right )}
     {\Lambda_B^2 (E_\gamma)} , 
\label{eq:B-to-ell-nu-gamma-DDW}
\end{equation}
where $x_\gamma = 2 E_\gamma/m_B$ is the reduced photon energy.    
The coefficient $C_3^{\rm SCET}$ results  
after matching QCD and SCET operators 
at the hard scale $\mu_F = m_B$:   
\begin{equation}
C_3^{\rm SCET} = 1 + \frac{\alpha_s (m_B) C_F}{4 \pi} \left [ 
- 2 \log^2 x_\gamma - 2 \, {\rm Li}_2 (1 - x_\gamma) + 
\frac{3 x_\gamma - 2}{x_\gamma - 1} \log x_\gamma - 6 - \frac{\pi^2}{12}  
\right ] . 
\label{eq:C3-SCET-NLO}
\end{equation}
The differential decay width~(\ref{eq:B-to-ell-nu-gamma-DDW}) 
depends on non-perturbative parameters~--- the first inverse 
moment of the leading DA $\phi^B_+ (k_+; \mu_F)$ and its 
logarithmic extensions in the form~\cite{DescotesGenon:2002mw}:  
\begin{equation}
\Lambda_B^{-1} (E_\gamma) = e^{- S (E_\gamma; \mu_F)} 
\int \frac{d k_+}{k_+} \, \phi^B_+ (k_+; \mu_F) \left \{ 
1 + \frac{\alpha_s C_F}{4 \pi} \left [ 
\log^2 \frac{2 E_\gamma k_+}{\mu_F^2} - \frac{\pi^2}{12} - 1 
\right ]  
\right \} . 
\label{eq:Lambda-B-inv-NLO}
\end{equation}
Here, the exponential factor $e^{- S (E_\gamma; \mu_F)}$ 
appears after resummation of large Sudakov logarithms.  
It should be noted that power corrections of order of 
${\cal O} (1/m_B, 1/(2 E_\gamma))$ to the decay rate 
are also calculated recently~\cite{Beneke:2011nf}.  

The $B^+ \to \ell^+ \nu_\ell \gamma$ decay was a subject 
of intensive experimental searches on the $B$-factories
at SLAC and KEK. In particular, the experimental analysis 
for the partial branching fraction in the photon-energy 
interval $E_\gamma^{\rm sig} = [1~{\rm GeV}, m_B/2]$ with 
the full dataset of $(771.6 \pm 10.6) \times 10^6$ $B \bar B$ 
pairs has been reported by the Belle collaboration 
recently~\cite{Heller:2015vvm}.   
%
The upper limits ($@ \, 90\%$~C.\,L.) presented are as follows:   
${\cal B} (B^+ \to e^+ \nu_e \gamma) < 6.1 \times 10^{-6}$, 
${\cal B} (B^+ \to \mu^+ \nu_\mu \gamma) < 3.4 \times 10^{-6}$, 
and ${\cal B} (B^+ \to \ell^+ \nu_\ell \gamma) < 3.5 \times 10^{-6}$. 
The last one was translated into the restriction 
on the first inverse moment~\cite{Heller:2015vvm}: 
\begin{equation}
\lambda_B = \left [ \int_0^\infty \frac{d k_+}{k_+} \, 
\phi^B_+ (k_+) \right ]^{-1} > 238~{\rm MeV} .  
\label{eq:Lambda-B-inv-Belle}
\end{equation}
This result is in good agreement with the theoretical 
estimates being within the interval~\cite{}: 
$300~{\rm MeV} < \lambda_B < 600~{\rm MeV}$.

The related topic is the shape of the $B^+ \to \gamma$ form factors 
entering the transition matrix element~(\ref{eq:B-to-gamma-TME}). 
Both the vector and axial-vector form factors with account 
of the soft contribution are known~\cite{Braun:2012kp}.   
%
Neglecting radiative and $1/m_B$ corrections, the form factors 
are equal $F_V^{(0)} = F_A^{(0)} = F_{B \to \gamma^*}^{(0)}$. 
The corresponding soft contribution is estimated by the LCSRs 
method~\cite{Braun:2012kp}:  
%
\begin{equation}
F_{B \to \gamma^*}^{(0)} (E_\gamma) = 
\frac{Q_u f_B m_B}{2 E_\gamma \lambda_B (\mu)} + 
\frac{Q_u f_B m_B}{2 E_\gamma} 
\int_0^{\omega_0} d\omega \left [ 
\frac{2 E_\gamma}{m_\rho^2} \, e^{- (2 E_\gamma \omega - m_\rho^2)/M^2} - 
\frac{1}{\omega}
\right ] \phi^B_+ (\omega, \mu) ,
\label{eq:F-B-gamma-soft}
\end{equation}
where $Q_u = 2/3$ is the electric charge of the $u$-quark,  
$M$ is the Borel parameter, and $m_\rho \simeq 775$~MeV 
is the $\rho$-meson mass~\cite{Olive:2016xmw}. 
Note that the isospin symmetry is implicitly assumed 
which means that the difference between the $\rho$- 
and $\omega$-mesons is neglected. 
The soft part of the form factor (the second term 
in~(\ref{eq:F-B-gamma-soft})) is also dependent 
on the other non-perturbative parameter~--- 
the effective threshold $s_0 = 1.2$~GeV, where 
$\omega_0 = s_0/(2 E_\gamma)$. It was also 
demonstrated that a choice of a model for the 
$B$-meson DA influences a result for the form 
factor but this dependence is not substantial. 

\subsection{$B_q \to \gamma \gamma$ and $B_q \to \ell^+ \ell^- \gamma$ Decays}  
\label{ssec:Parkh-B-to-2gamma-and-ell-ell-decays}

The interesting problem within the SCET is the universality 
of non-perturbative effects in leptonic and radiative $B$-meson decays. 
Let us start with the radiative $B_q \to \gamma \gamma$ decay, where $q = d$ or~$s$, 
the decay width of which can be written as follows~\cite{DescotesGenon:2002ja}:    
%
\begin{equation}
\Gamma = \frac{\alpha^2 G_F^2 f_B^2 m_B^5}{144 \pi^3} \, 
|V_{tq} V_{tb}^*|^2 
\left | C_7^{\rm eff} C_9^{\rm SCET} \right |^2
\frac{1}{\Lambda_B^2 (m_B/2)} . 
\label{eq:B-2gamma-width}
\end{equation}
Similar processes are the ones where one of the photons 
is virtual and decaying into the lepton pair. 
The differential decay width of $B_q \to \ell^+ \ell^- \gamma$, 
where $\ell = e, \, \mu$, has the form~\cite{DescotesGenon:2002ja}: 
%
\begin{eqnarray}
\frac{d\Gamma}{dE_\gamma} =  
\frac{\alpha^3 G_F^2 f_B^2 m_B^4}{1728 \pi^4} \, 
|V_{tq} V_{tb}^*|^2 
\frac{x_\gamma \left ( 1 - x_\gamma \right )}
     {\Lambda_B^2 (E_\gamma)} 
\left [ 
\left | C_9^{\rm eff} C_3^{\rm SCET} + 
        \frac{2 C_7^{\rm eff}}{1 - x_\gamma} \, C_9^{\rm SCET} 
\right |^2 + \left | C_{10} C_3^{\rm SCET} \right |^2 
\right ] , 
\label{eq:B-ell-ell-gamma-width}  
\end{eqnarray}
where $x_\gamma = 2 E_\gamma/m_B$. 
The coefficient $C_3^{\rm SCET}$~(\ref{eq:C3-SCET-NLO}) determines 
the differential width of the $B^+ \to \ell^+ \nu_\ell \gamma$
decay while the decays considered above are dependent on the other 
coefficient $C_9^{\rm SCET}$ which also results after perturbative 
matching of QCD and SCET operators at the scale $\mu_F = m_B$: 
\begin{eqnarray}
C_9^{\rm SCET} = 1 + \frac{\alpha_s (m_B) C_F}{4 \pi} \left [ 
\log \frac{m_B^2}{\mu_R} - 2 \log^2 x_\gamma + 2 \log x_\gamma - 
2 \, {\rm Li}_2 (1 - x_\gamma) - 6 - \frac{\pi^2}{12}  
\right ] . 
\label{eq:C9-SCET-NLO}  
\end{eqnarray}

These examples explicitly show the necessity to know two coefficients 
$C_3^{\rm SCET}$ and $C_9^{\rm SCET}$ only for performing theoretical 
analysis of the radiative and leptonic radiative $B$-mesons decays. 
As for the $B$-meson DAs, all three decays considered are dependent 
on $\Lambda_B^{-1} (\mu)$~(\ref{eq:Lambda-B-inv-NLO}) and this quantity 
can be independently determined from the analysis of each decay. 
This can be a good universality test of the soft contribution  
determined by the $B$-meson dynamics.

\subsection{$B^+ \to \pi^+ \ell^+ \ell^-$ Decays}  
\label{ssec:Parkh-B-to-pi-ell-ell-decays}

The other good example of the $B$-meson decays which are 
sensitive to the $B$-meson distribution amplitudes are rare 
semileptonic decays like the $B^+ \to P \, \ell^+ \ell^-$ decay, 
where $P = \pi,\, K,\, \eta^{(\prime)}$ and $\ell = e,\, \mu$. 
In this lecture some details of the $B^+ \to \pi^+ \ell^+ \ell^-$ 
decay are presented. 

Detailed perturbative analysis in full kinematical region 
of~$q^2$, the lepton-pair invariant mass squared, was 
undertaken in~\cite{Ali:2013zfa}.   
%
The differential branching fraction is as follows: 
\begin{equation}
\frac{d \mbox{Br} \left ( B^+ \to \pi^+ \ell^+ \ell^- \right )}{d q^2} = 
\frac{G_F^2 \alpha_{\rm em}^2 \tau_B}{1024 \pi^5 m_B^3} 
|V_{tb} V_{td}^*|^2  
\sqrt{\lambda (q^2)} \sqrt{1-\frac{4 m_\ell^2}{q^2}} 
F (q^2) .  
\label{eq:B-to-pi-ell-ell-Diff-BF}
\end{equation}
This expression contains the dynamical function: 
\begin{eqnarray}
F (q^2) & = & \frac{2}{3} \, \lambda (q^2) 
\left ( 1 + \frac{2 m_\ell^2}{q^2} \right )
\left | C_9^{\rm eff} f_+ (q^2) + 
\frac{2 m_b}{m_B + m_\pi} \, C_7^{\rm eff} f_T (q^2) \right |^2 
\label{eq:B-to-pi-ell-ell-Dyn-F} \\ 
& + & \frac{2}{3} \, \lambda(q^2) 
\left ( 1 - \frac{4 m_\ell^2}{q^2} \right ) 
\left | C_{10}^{\rm eff} \right |^2 f_+^2 (q^2) + 
\frac{4 m_\ell^2}{q^2} \left ( m_B^2 - m_\pi^2 \right)^2 
\left | C_{10}^{\rm eff} \right |^2 f_0^2 (q^2) ,  
\nonumber 
\end{eqnarray}
where $C_i^{\rm eff}$ are the effective Wilson coefficients  
which are specific combinations of Wilson coefficients 
entering the effective weak $b \to d$ Hamiltonian: 
\begin{equation}
{\cal H}_{\rm eff}^{(b \to d)} = 
- \frac{4 G_F}{\sqrt 2} \! \biggl [ V_{tb}^* V_{td} 
\sum\limits_{i=1}^{10} \! C_i (\mu) {\cal O}_i (\mu) + 
V_{ub}^* V_{ud} \sum \limits_{i=1}^{2} \! 
C_i (\mu) \! \left ( \! 
{\cal O}_i (\mu) - {\cal O}_i^{(u)} (\mu) \! 
\right) \! \biggr ]  + {\rm h.\,c.}  
\label{eq:eff-Hamiltonian}
\end{equation}
Here, $G_F$ is the Fermi constant,   
$C_i (\mu)$ are Wilson coefficients,  
${\cal O}_i (\mu)$ are the dimension-six operators, and  
$V_{ij}$ are Cabibbo-Kobayashi-Maskawa (CKM) matrix elements. 
One can easily recognize that the products 
$V_{tb}^* V_{td} \sim V_{ub}^* V_{ud} \sim \lambda^3$ 
are of the same order in $\lambda = \sin\theta_C$, 
where $\theta_C$ is the Cabibbo angle. 

The dynamical function~(\ref{eq:B-to-pi-ell-ell-Dyn-F}) 
is depended on three form factors $f_+ (q^2)$, $f_0 (q^2)$, 
and $f_T (q^2)$ which are non-perturbative scalar functions 
of the momentum transfered squared. They are entering the  
vector and tensor $B \to \pi$ transition matrix elements. 
The Heavy-Quark Symmetry (HQS) is applicable in the 
large-recoil limit (small $q^2$-values) and relates these 
form factors~\cite{Beneke:2000wa}: 
%
%
\begin{eqnarray}
f_0 (q^2) & = & \left ( \frac{m_B^2 + m_\pi^2 - q^2}{m_B^2} \right )
\Biggl [ \left \{ 1 + \frac{\alpha_s (\mu) C_F}{4\pi} 
\left ( 2 - 2 L (q^2) \right ) \right \} f_+ (q^2) 
\nonumber \\ 
& + & \frac{\alpha_s (\mu) C_F}{4\pi} 
\frac{m_B^2 (q^2 - m_\pi^2)}{(m_B^2 + m_\pi^2 - q^2)^2}
\Delta F_\pi \Biggr ] , 
\label{eq:f0-fp-relation} \\
f_T (q^2) & = & \left ( \frac{m_B + m_\pi}{m_B} \right )
\Biggl [ \left ( 1 + \frac{\alpha_s (\mu) C_F}{4\pi} 
\left ( \ln \frac{m_b^2}{\mu^2} + 2 L (q^2) \right ) \right ) f_+ (q^2) 
\nonumber \\ 
& - & \frac{\alpha_s (\mu) C_F}{4\pi} \frac{m_B^2}{m_B^2 + m_\pi^2 - q^2}
\Delta F_\pi \Biggr ] , 
\label{eq:fT-fp-relation} 
\end{eqnarray}
where for simplicity the following quantities are introduced:  
\begin{equation}
L (q^2) = 
\left ( 1 + \frac{m_B^2}{m_\pi^2 - q^2} \right )
\ln \! \left ( 1 + \frac{m_\pi^2 - q^2}{m_B^2} \right ) ,  
\qquad 
\Delta F_\pi = \frac{8 \pi^2 f_B f_\pi}{N_c m_B \lambda_B} 
\left < \bar u^{-1} \right >_\pi . 
\label{eq:Lq2-DeltaF-def}
\end{equation}
The last quantity $\Delta F_\pi$ contains 
the first inverse moments of $\pi$- and $B$-meson:
\begin{equation}
\left < \bar u^{-1} \right >_\pi = 
\int \! du \frac{\phi_\pi (u)}{1 - u}, 
\qquad 
\lambda_B^{-1} \equiv \left < \bar \omega^{-1} \right >_+ = 
\int \! d\omega \frac{\phi^B_+ (\omega)}{\omega} .
\label{eq:u-inv-omega-plus-inv-def}
\end{equation}
Only one form factor $f_+ (q^2)$ is required 
for getting the $q^2$-distribution in this decay 
which can be fitted from the data on the 
$B \to \pi \ell^+ \nu_\ell$ decays~\cite{Ali:2013zfa}. 

Within the factorization approach, it is also possible 
to calculate some types of power-suppressed corrections 
in the $B \to \pi \ell^+ \ell^-$ decay, in particular, 
the annihilation contributions. 
This type of corrections contains the $q^2$-dependent 
first inverse moment of the sub-leading $B$-meson LCDA:  
\begin{equation}
\lambda_{B,-}^{-1} (q^2) = \int_0^\infty \!\!
\frac{\phi^B_- (\omega) \, d\omega}{\omega - q^2/M_B - i \epsilon} . 
\label{eq:q2-omega-minus-inv-def}
\end{equation}
Note the specific feature of this moment: it is logarithmically 
divergent at $q^2 \to 0$ because the sub-leading LCDA 
$\phi^B_- (\omega)$ turns out to be constant at small~$\omega$,   
$\phi^B_- (\omega) |_{\omega \to 0} \sim {\rm const}$. 
Nevertheless, such a fiture does not result a problem 
in the numerical analysis of the differential branching 
fraction as the kinematical lower cut $q^2 \ge 4 m_\ell^2$ 
exists in the semileptonic $B \to \pi \ell^+ \ell^-$ decay.  
Annihilation contributions of a similar type enter 
also in decay amplitudes of similar semileptonic 
$B \to V_\| \ell^+ \ell^-$ decays, where~$V_\|$ 
is the longitudinally polarized light vector 
meson~\cite{Beneke:2004dp}.  
Decays with a transversely polarized vector meson 
in the final state are dependent on $\phi^B_+ (\omega)$ 
to the leading order. In the limit $q^2 = 0$, the 
corresponding branching fraction is finite and can 
be related with the branching ratios of rare radiative 
$B \to V \gamma$ decays.

\subsection{Models for the $B$-Meson Distribution Amplitudes}
\label{ssec:Parkh-Models} 

\begin{figure}[tb]
\centerline{
\includegraphics[width=0.45\textwidth]{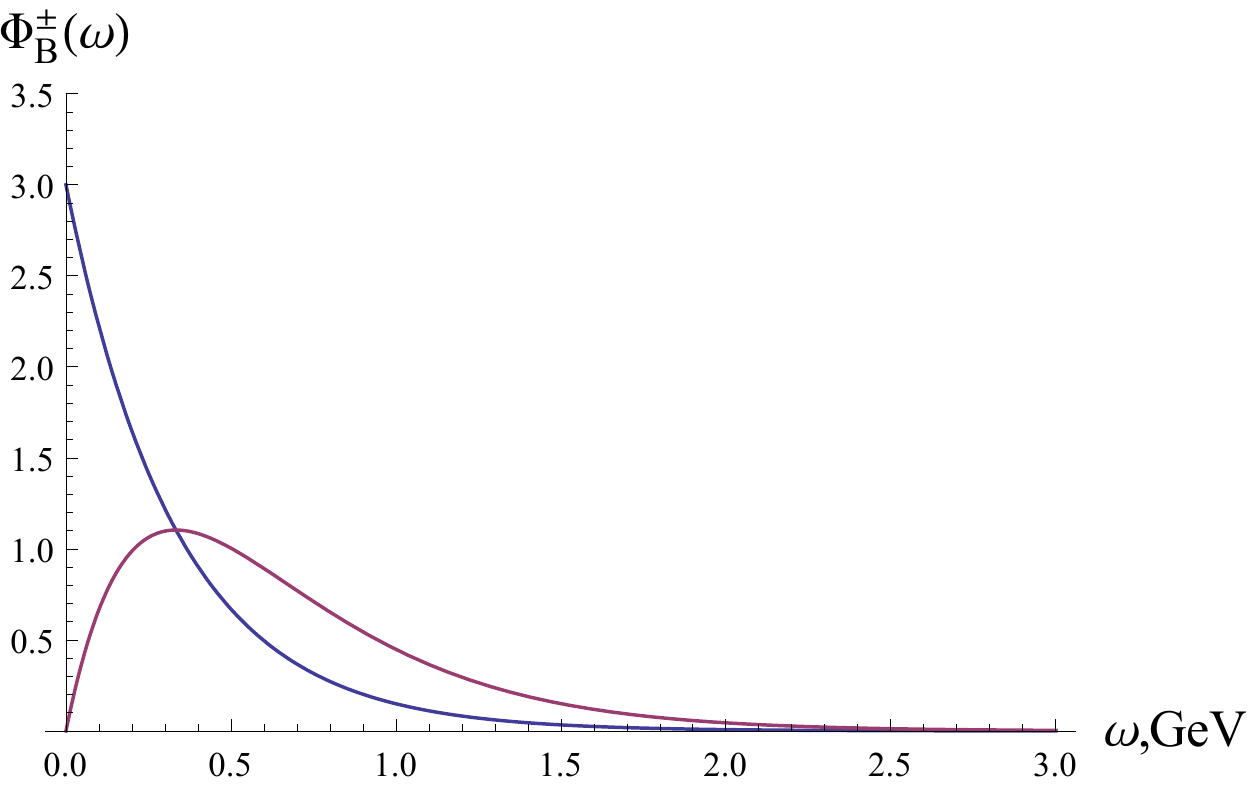} 
\hfill 
\includegraphics[width=0.45\textwidth]{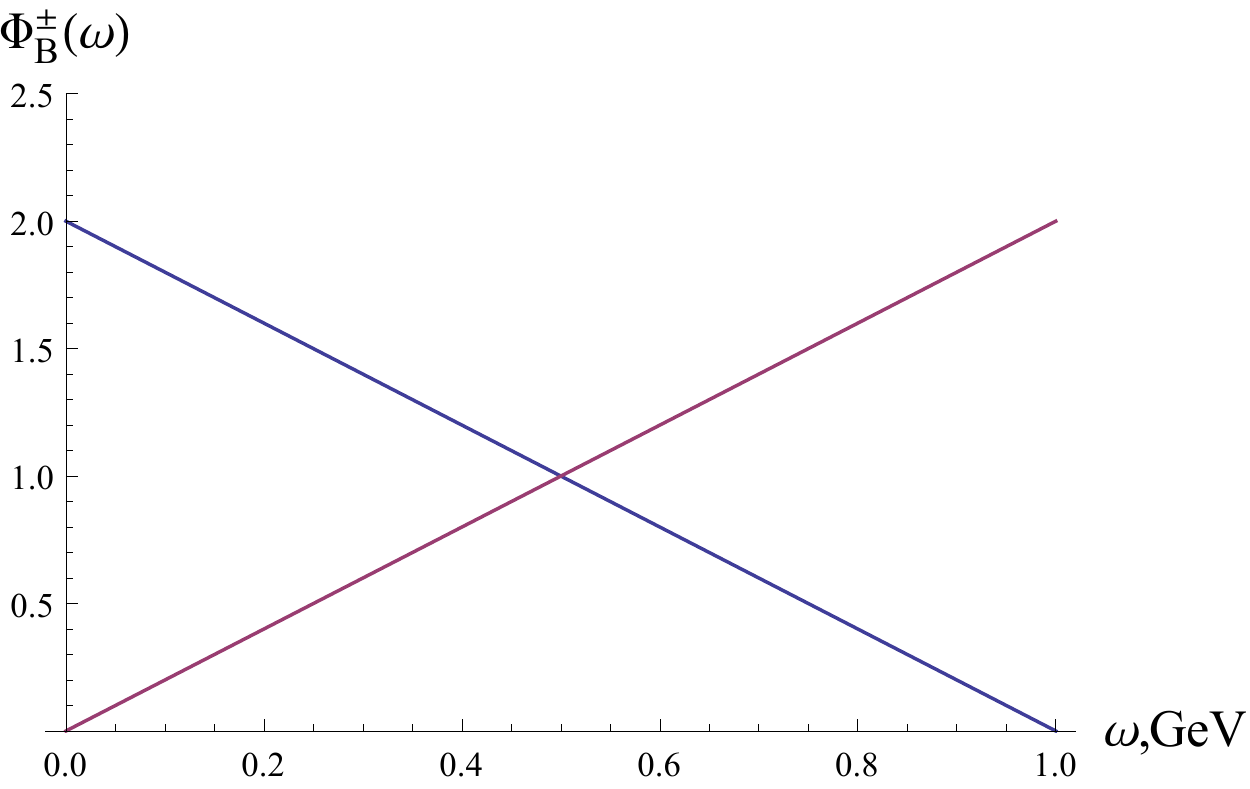} 
}		
\caption{
The energy dependence of of both leading and non-leading 
distribution amplitude models where the exponential ones 
are presented on the left panel and the light-meson-like 
models are on the right panel.
}
\label{fig:B-Meson-DAs-models}
\end{figure}

The distribution amplitudes are non-perturbative quantities 
and usual perturbative methods of QFT~\cite{Peskin:1995ev} are 
inapplicable. The commonly used method of their calculation 
is the QCD Sum Rules~\cite{Grozin:1996pq,Braun:2003wx}. 
To combine these distribution amplitudes with hard kernels 
in amplitudes of physical processes, one needs to model them 
by some analytical expressions called the distribution amplitude 
models. At present, several models for the distribution amplitudes 
have been suggested. Two simplest ones are the exponential 
models~\cite{Grozin:1996pq}:  
\begin{equation}
\phi^+_B (\omega) = \frac{\omega}{\omega_0^2} \, e^{-\omega/\omega_0},
\qquad 
\phi^-_B (\omega) = \frac{1}{\omega_0} \, e^{-\omega/\omega_0} ,
\label{eq:GN-models}
\end{equation}
where $\omega_0 = 2\bar\Lambda/3$, which appeared 
to be the most popular in physical applications, 
and the light-meson-like models~\cite{Kawamura:2001jm}:  
\begin{equation}
\phi^+_B (\omega) = \frac{\omega}{2 \bar\Lambda^2} \, 
\theta (2 \bar\Lambda - \omega) ,
\qquad 
\phi^-_B (\omega) = \frac{2\bar\Lambda-\omega}{2\bar\Lambda^2} \, 
\theta (2 \bar\Lambda - \omega) , 
\label{eq:KKQT-models}
\end{equation}
which is also simple and explicitly based 
on the light-meson distribution amplitude. 
The energy dependence of these distribution amplitudes 
is presented in Fig.~\ref{fig:B-Meson-DAs-models}. 

More involved models are the two-parametric model by Braun, Ivanov, 
and Korchemsky (BIK)~\cite{Braun:2003wx} and improved exponential 
model by~Lee and Neubert~\cite{Lee:2005gza}. The later one matches 
the exponential behavior at low momenta of light quark and QCD-based 
behavior (the radiative tail) at large momenta. For these models 
the leading distribution amplitude only was studied while the 
non-leading amplitude was skipped. 

As mentioned above, one needs moments of these amplitudes 
in getting decay widths of heavy mesons. Among moments, 
first inverse ones are of special interest. 
The uncertainty induced by a choice of the distribution 
amplitude model of the heavy meson on the semileptonic 
decay rates of $B$-mesons is also interesting to study
which is not yet been worked out.

\section{Light-Cone Distribution Amplitudes of Heavy Baryons} 
\label{sec:Parkh-LCDAs-b-Baryons}

Light-cone distribution amplitudes (LCDAs) 
of heavy baryons are the transition matrix 
elements from the baryonic state to vacuum 
of non-local light-ray operators built off 
an effective heavy quark and two light quarks.  
The content of such operators supports 
a similarity in the construction of the 
heavy-baryon LCDAs to both the $B$-meson 
(within the HQET)~\cite{Grozin:1996pq,Braun:2003wx} 
and the nucleon (within QCD)~\cite{Braun:1999te,Braun:2000kw} 
LCDAs descriptions.   
An important simplifying feature of the 
operators containing one or more heavy 
quarks is an existence of the Heavy Quark 
Symmetry (HQS) which results into the 
decoupling of the heavy-quark spin 
from the system dynamics in the limit 
$m_Q \to \infty$, 
where~$m_Q$ is the heavy-quark mass.   
So, to understand the properties of heavy 
baryons in this limit, it is enough to 
switch off the heavy-quark spin and to 
introduce a total set of two-particle 
LCDAs corresponding to the light-quark 
system, called diquark, which quantum 
numbers completely determine a number 
of LCDAs and their asymptotic behavior.  
 
In this simplified picture, there are 
the $SU(3)_F$ antitriplet of ``scalar 
baryons'' with the $J^P = 0^+$ spin-parity 
determined by the diquark spin-parity 
$j^p = 0^+$ and the $SU(3)_F$ sextet 
of ``axial-vector baryons'' with the 
$J^P = 1^+$ spin-parity which follows 
from the diquark spin-parity $j^p = 1^+$. 
It is reasonable to start with the 
description of the ``scalar baryons'' 
and then to generalize the procedure 
on the ``axial-vector baryons''. 
The changes originated by an account 
of the heavy-quark spin can be done 
after the total sets of the non-local 
operators and corresponding LCDAs  
are introduced in the decoupling limit. 
All these steps are discussed briefly 
in this section.

\subsection{``Scalar Baryons''}  

The ``scalar baryons'' are combined into 
the $SU(3)_F$ antitriplet with $J^P = 0^+$  
in which the light diquark states are also 
the scalar states with $j^p = 0^+$. 

The set of the LCDAs is determined by the matrix 
elements between the baryonic state and vacuum 
of the four independent non-local light-ray 
operators~\cite{Ball:2008fw,Ali:2012zza,Ali:2012pn}:  
\begin{eqnarray}
 \epsilon^{abc} \langle 0| \left ( 
 q_1^a (t_1 n) C \gamma_5 \hat n q_2^b (t_2 n) 
 \right ) h^c_v (0) |H (v) \rangle & = & 
 f^{(2)}_H \, \Psi_2 (t_1,t_2) , 
\label{eq:def-LCDAs} \\
 \epsilon^{abc} \langle 0| \left ( 
 q_1^a (t_1 n) C \gamma_5 q_2^b (t_2 n) 
 \right ) h^c_v (0) |H (v) \rangle & = & 
 f^{(1)}_H \, \Psi_3^s (t_1,t_2) , 
\nonumber\\ 
 \epsilon^{abc} \langle 0| \left ( 
 q_1^a (t_1 n) C \gamma_5 i \sigma_{\bar n n} q_2^b (t_2 n) 
 \right ) h^c_v (0) |H (v) \rangle & = & 
 2 f^{(1)}_H \, \Psi_3^\sigma (t_1,t_2) , 
\nonumber\\ 
 \epsilon^{abc} \langle 0| \left ( 
 q_1^a (t_1 n) C \gamma_5 \hat{\bar n} q_2^b (t_2 n) 
 \right ) h^c_v (0) |H (v) \rangle & = & 
 f^{(2)}_H \Psi_4 (t_1,t_2) , 
\nonumber 
\end{eqnarray}
where $q_i (x) = u (x),\, d(x),\, s(x)$ 
are the light-quark fields, $h_v (0)$ 
is the static heavy-quark field situated 
at the origin of the position-space frame,  
$C$ is the charge conjugation matrix, 
$n^\mu$ and $\bar n^\mu$ are 
two light-like vectors normalized 
by the condition $(n \bar n) = 2$
\footnote{
The definitions of~$n^\mu$ and $\bar n^\mu$ 
differ by the factor~$1/\sqrt 2$ from 
the ones used in the previous section. 
}, 
and 
$\sigma_{\bar n n} = i \left ( 
\hat{\bar n} \hat n - \hat n \hat{\bar n}   
\right )/2$.   
In addition, the frame is adopted where 
the heavy-meson velocity is related 
to the light-like vectors as follows: 
$v^\mu = \left ( n^\mu + \bar n^\mu \right )/2$.   
The light-quark fields  
on the light cone are assumed 
to be multiplied by the Wilson lines:   
$$
q (t n) = \left [ 0, t n \right ] q (t n) = 
{\rm P} \exp \left \{ - i g_{\rm st} t 
\int_0^1 d\alpha n^\mu \, A_\mu (\alpha t n) 
\right \} q (t n) = \sum_{N = 0}^\infty \frac{t^N}{N!} \, 
\left ( n^\mu D_\mu \right )^N q (0) ,  
$$  
where the following definition 
of the covariant derivative   
$D_\mu = \partial_\mu - i g_{\rm st} A_\mu$ 
is accepted. 
%
%
The static heavy-quark field living on the light cone 
also includes the Wilson line but of the other type 
with the time-like link~\cite{Korchemsky:1991zp}:  
$$
h_v (0) = {\rm P} \exp \left \{ i g_{\rm st}  
\int_{-\infty}^0 d\alpha v^\mu \, A_\mu (\alpha v) 
\right \} \phi (- \infty) ,  
$$  
with which it is connected with the sterile field $\phi (- \infty)$.

The couplings~$f^{(i)}_H$ introduced in Eqs.~\eqref{eq:def-LCDAs} 
to make the LCDAs dimensionless are defined by local operators%
~\cite{Grozin:1992td,Bagan:1993ii,Groote:1996em,Groote:1997yr}:  
\begin{eqnarray}
&& 
  \epsilon^{abc} \langle 0| \left ( 
  q_1^a (0) C \gamma_5  q_2^b(0) 
  \right ) h^c_v (0) |H(v) \rangle = f^{(1)}_H , 
\nonumber\\
&& 
 \epsilon^{abc} \langle 0| \left ( 
 q_1^a (0) C \gamma_5  \hat v q_2^b (0) 
 \right) h^c_v (0) |H (v) \rangle = f^{(2)}_H . 
\nonumber 
\end{eqnarray} 
The scale dependences of these couplings are governed 
by the anomalous dimensions~$\gamma^{(i)}$ of local 
operators as follows:  
$$ 
\frac{d\ln f_H^{(i)}(\mu)}{d\ln \mu}
\equiv - \gamma^{(i)} 
= - \sum_k \gamma^{(i)}_k \, a^k (\mu), 
\qquad 
a (\mu) \equiv \frac{\alpha_s^{\overline{\rm MS}}(\mu)}{4\pi},  
$$ 
where the strong coupling is determined 
in the $\overline{\rm MS}$-scheme. 
This equation can be solved order by order 
in the $a (\mu)$-power expansion and in the 
NLO order, one can use the following formula: 
\begin{eqnarray}
 f^{(i)}_H (\mu) = f^{(i)}_H (\mu_0) \left ( 
 \frac{\alpha_s (\mu)}{\alpha_s (\mu_0)} 
 \right )^{\gamma^{(i)}_1/\beta_0} \left [ 
 1 - \frac{\alpha_s (\mu_0) - \alpha_s(\mu)}{4\pi} \, 
 \frac{\gamma^{(i)}_1}{\beta_0} \left ( 
 \frac{\gamma^{(i)}_2}{\gamma^{(i)}_1} - \frac{\beta_1}{\beta_0} 
 \right ) \right ] , 
\nonumber  
\end{eqnarray}
where $\beta_{0,1}$ are the first two coefficient 
in the perturbative expansion of the $\beta$-function.  
As the evolution to the required scale can be easily 
done now, one needs to know numerical values of the 
couplings~$f^{(i)}_H (\mu)$ at some representative 
scale~$\mu_0$, say $\mu_0 = 1$~GeV. As this scale 
is rather low to use the perturbation theory, 
non-perturbative techniques are necessary to 
calculate the value. In particular, the QCD sum rules method 
in NLO for the $\Lambda_b$-baryon results~\cite{Groote:1997yr}:   
%
$$
f^{(1)}_{\Lambda_b} (\mu_0 = 1~{\rm GeV}) \simeq 
f^{(2)}_{\Lambda_b} (\mu_0 = 1~{\rm GeV}) \simeq  
0.030 \pm 0.005~\mbox{\rm GeV}^3 .   
$$ 
%
%
%
Non-relativistic constituent-quark picture 
of heavy baryons~$H$  
suggests that $f^{(2)}_H \simeq f^{(1)}_H$ 
at low scales of order 1~GeV, and this expectation 
is supported by numerous QCD sum rule calculations%
~\cite{Bagan:1993ii,Grozin:1992td,Groote:1996em,Groote:1997yr}.
These couplings~$f^{(i)}_H (\mu)$ cannot coincide 
at all scales because of different anomalous 
dimensions~$\gamma^{(i)}$ of local operators. 

Similar to the couplings~$f^{(i)}_H (\mu)$, 
the LCDAs $\Psi_i (t_1, t_2)$ introduced 
in Eq.~\eqref{eq:def-LCDAs} are also 
scale-dependent functions.   
To find their scale evolution, it is convenient to 
make their Fourier transform to the momentum space:  
\begin{eqnarray}
  \Psi (t_1,t_2) =  
  \int_0^\infty \!\!d\omega_1  \int_0^\infty\!\! d\omega_2 \, 
  e^{-i t_1 \omega_1 -i t_2 \omega_2} \psi (\omega_1, \omega_2) 
 = \int_0^\infty \!\!\omega\, d\omega \int_0^1 du \, 
 e^{-i \omega (t_1 u + t_2 \bar u)} \, \widetilde\psi (\omega, u) , 
\nonumber 
\end{eqnarray}
%
%
where $\bar u = 1 - u$. 
In the first representation 
$\omega_1 = u \omega$ and  
$\omega_2 = \left ( 1 - u \right ) \omega = \bar u \omega$ 
are the energies of the light quarks~$q_1$ and~$q_2$.   
The leading-order evolution equation 
for $\psi_2 (\omega_1, \omega_2; \mu)$     
can be derived by identifying the ultra-violet 
singularities of the one-gluon-exchange 
diagrams~\cite{Ball:2008fw}.  
%
The evolution equation in the leading order 
is expressed in terms of two-particle kernels:  
%
\begin{eqnarray}
&& 
  \mu \frac{d}{d\mu} \, \psi_2 (\omega_1, \omega_2; \mu) =  
  - \frac{\alpha_s (\mu)}{2\pi} \, \frac{4}{3} \, \Bigg \{
  \int_0^\infty d\omega_1' \, 
  \gamma^{\rm LN} (\omega_1', \omega_1; \mu) \, 
  \psi_2 (\omega'_1, \omega_2; \mu)
\nonumber \\
&& \hspace*{3mm}
  + \int_0^\infty d\omega_2' \, 
  \gamma^{\rm LN} (\omega_2', \omega_2; \mu) \, 
  \psi_2 (\omega_1, \omega'_2; \mu)
   - \int_0^1 dv \, V (u, v) \, \psi_2 (v \omega, \bar v \omega; \mu) + 
   \frac{3}{2} \, \psi_2 (\omega_1, \omega_2; \mu) \Bigg\} , 
\nonumber 
\end{eqnarray}
where the kernel $\gamma^{\rm LN} (\omega', \omega; \mu)$ 
controls the evolution of the $B$-meson LCDA~\cite{Lange:2003ff} 
and~$V (u, v)$ is the Efremov-Radyushkin-Brodsky-Lepage (ER-BL) 
kernel~\cite{Efremov:1978rn,Lepage:1979zb}.  
The term $3 \psi_2 (\omega_1, \omega_2; \mu) / 2$ results 
from the one-loop $f^{(2)}_H$ renormalization subtraction.   
The evolution equation above can be solved either numerically 
or semi-analytically~\cite{Ball:2008fw,Ali:2012pn}.   

The next step in working out solutions of the heavy-baryon 
evolution equations analytically was undertaken in~\cite{Bell:2013tfa}. 
In particular, the eigenfunctions of the Lange-Neubert 
evolution kernel were found and used for a systematic 
implementation of the renormalization-group effects for both 
the $B$-meson and $\Lambda_b$-baryon wave-function evolutions. 
Based on these foundations, the new strategy to construct 
the LCDA models in accordance with the Wandzura-Wilczek-like 
relations was presented. As a possible extension of the above 
analysis in application to baryons, the classification 
of the non-local baryonic operators constructed from four 
particles (three quarks and a gluon) is required to work out 
equations involving explicitly the there-particle LCDAs 
and twist-four four-particle ones which should reduce to the 
Wandzura-Wilczek relations after four-particle LCDAs are neglected.

\subsection{``Axial-Vector Baryons''}  

The ``axial-vector baryons'' are components 
of the $SU(3)_F$ sextet with $J^P = 1^+$  
in which the light diquark states are  
the axial-vector states with $j^p = 1^+$. 
In a difference to the ``scalar baryons'' 
case, one needs to consider the baryons 
with the longitudinal and transverse 
polarizations separately.  

The set of the longitudinal LCDAs is determined 
by the matrix elements between the baryonic state 
with the appropriate polarization and vacuum 
of the four independent non-local light-ray 
operators~\cite{Ali:2012zza,Ali:2012pn}:    
\begin{eqnarray} 
\epsilon^{abc} \langle 0| \left ( 
q_1^a (t_1) C \, \hat n q_2^b (t_2) 
\right )  h_v^c (0) | H (v, \varepsilon) \rangle  
& = & \left ( \bar v \varepsilon \right ) 
f^{(2)}_H \, \Psi_2^\| (t_1, t_2) 
\nonumber \\
\epsilon^{abc} \langle 0| \left ( 
q_1^a (t_1) C \, q_2^b (t_2) 
\right )  h_v^c (0) | H (v, \varepsilon) \rangle  
& = & \left ( \bar v \varepsilon \right ) 
f^{(1)}_H \, \Psi_3^{\| s} (t_1,t_2) 
\nonumber \\
\epsilon^{abc} \langle 0| \left ( 
q_1^a (t_1) C \, i \sigma_{\bar n n} q_2^b (t_2)   
\right )  h_v^c (0) | H (v, \varepsilon) \rangle  
& = & 2 \left ( \bar v \varepsilon \right ) 
f^{(1)}_H \, \Psi_3^{\| a} (t_1,t_2) 
\nonumber \\
\epsilon^{abc} \langle 0| \left ( 
q_1^a (t_1) C \, \hat{\bar n} q_2^b (t_2) 
\right )  h_v^c (0) | H (v, \varepsilon) \rangle  
& = & - \left ( \bar v \varepsilon \right ) 
f^{(2)}_H \, \Psi_4^\| (t_1, t_2) 
\nonumber 
\end{eqnarray}
where 
$\bar v^\mu = \left ( \bar n^\mu - n^\mu \right )/2$ 
is the four-vector normalized as $(\bar v \bar v) = -1$ 
and orthogonal to the four-velocity $(v \bar v) = 0$. 
In the LCDA definitions above, the baryonic state 
is assumed to have a pure longitudinal 
polarization~$\varepsilon_\|^\mu = \bar v^\mu$  
and the prefactor on the r.h.s. is simply   
$(\bar v \varepsilon) = -1$.   

The similar set of the transverse LCDAs is determined 
as follows~\cite{Ali:2012zza,Ali:2012pn}:    
\begin{eqnarray} 
\hspace*{-5mm} 
\epsilon^{abc} \langle 0| \left ( 
q_1^a (t_1) C \, \gamma_\perp^\mu \hat n q_2^b (t_2) 
\right )  h_v^c (0) | H (v, \varepsilon) \rangle  
& = & f^{(2)}_H \, \Psi_2^\perp (t_1, t_2) \, \varepsilon_\perp^\mu  
\nonumber \\
\hspace*{-5mm} 
\epsilon^{abc} \langle 0| \left ( 
q_1^a (t_1) C \, \gamma_\perp^\mu q_2^b (t_2) 
\right )  h_v^c (0) | H (v, \varepsilon) \rangle  
& = & f^{(1)}_H \, \Psi_3^{\perp s} (t_1,t_2) \, \varepsilon_\perp^\mu 
\nonumber \\
\hspace*{-5mm} 
\epsilon^{abc} \langle 0| \left ( 
q_1^a (t_1) C \, \gamma_\perp^\mu i \sigma_{\bar n n} q_2^b (t_2)   
\right )  h_v^c (0) | H (v, \varepsilon) \rangle  
& = & 2 f^{(1)}_H \, \Psi_3^{\perp a} (t_1,t_2) \, \varepsilon_\perp^\mu 
\nonumber \\
\hspace*{-5mm} 
\epsilon^{abc} \langle 0| \left ( 
q_1^a (t_1) C \, \gamma_\perp^\mu \hat{\bar n} q_2^b (t_2) 
\right )  h_v^c (0) | H (v, \varepsilon) \rangle  
& = & f^{(2)}_H \, \Psi_4^\perp (t_1, t_2) \, \varepsilon_\perp^\mu 
\nonumber 
\end{eqnarray}
where $\gamma_\perp^\mu = \gamma^\mu - \left ( 
\hat{\bar n} \hat n + \hat n \hat{\bar n}   
\right )/2$ and 
$\varepsilon_\perp^\mu = \varepsilon^\mu - \varepsilon_\|^\mu$
is the transverse polarization of the baryon.

\subsection{Real Baryons}
\label{ssec:Parkh-real}

As far as all the sets of the LCDAs are determined, 
it necessary to generalize their definitions to real 
baryons which simply means that the spin of the heavy 
quark should be included into the baryon wave function.    
In other words, the r.\,h.\,s. of matrix elements 
of all non-local operators must be multiplied on 
the Dirac spinor~$U (v)$ of the heavy quark~$h_v$, 
satisfying the conditions: $\hat v \, U (v) = U (v)$  
and $\overline U (v) \, U (v) = 1$.   
After these modifications, the ``scalar baryons'' 
transform to the baryons with the spin-parity 
$J^P = 1/2^+$ and the heavy-quark Dirac spinor~$U (v)$ 
is nothing else but the heavy-baryon spinor~$H (v)$, 
i.\,e. the spin of the heavy quark completely 
determines the spin structure of the heavy-baryon 
wave function. The case of ``axial-vector baryons'' 
is a little bit more complicated. It is well-known 
from quantum mechanics that the direct product 
of two angular momenta $j_1 = 1/2$ and $j_2 = 1$ 
is decomposed into two irreducible representations 
with the momenta $J_1 = 1/2$ and $J_2 = 3/2$.   
That is exactly the situation after the 
heavy-quark spin is switched on in the heavy 
baryon with the diquark in the axial-vector 
state $j^p = 1^+$:  
\begin{eqnarray} 
\varepsilon_\mu \, U (v) =  
\left [ 
\varepsilon_\mu \, U (v) - \frac{1}{3}  
\left ( \gamma_\mu + v_\mu \right ) 
\hat\varepsilon \, U (v) 
\right ] + \frac{1}{3}  
\left ( \gamma_\mu + v_\mu \right ) 
\hat\varepsilon \, U (v) 
\equiv  
R^{3/2}_\mu (v) + \frac{1}{3}  
\left ( \gamma_\mu + v_\mu \right ) H (v) .   
\nonumber 
\end{eqnarray}
As the result, there are two states with the 
spin-parities $J^P = 1/2^+$ and $J^P = 3/2^+$. 
The former one is described by the Dirac spinor~$H (v)$ 
and for the $J^P = 3/2^+$ state the Rarita-Schwinger 
vector-spinor $R^{3/2}_\mu (v)$, which satisfies 
the relations $\hat v \, R^{3/2}_\mu (v) = R^{3/2}_\mu (v)$, 
$v^\mu \, R^{3/2}_\mu (v) = 0$, and
$\gamma^\mu \, R^{3/2}_\mu (v) = 0$,    
is introduced.

\subsection{QCD Sum Rules} 
\label{ssec:Parkh-QCD-SRs}  

In applications to a calculation of amplitudes  
with heavy baryons, one needs to know realistic 
models for LCDAs. Such models can be obtained 
by matching several few moments of LCDA models 
and the corresponding ones calculated by some 
non-perturbative methods, say by the QCD sum rules. 
The later method requires a calculation of a  
two-point correlator 
which involve the non-local light-ray operator 
and a suitable local current $J^{\Gamma'} (x)$.  
%
The general structure of the heavy-baryon 
local current can be chosen as follows:  
\begin{equation*}
\bar J^{\Gamma'} (x) = \epsilon^{abc} \left ( \bar q_2^a (x) 
\left [ A + B \, \hat v \right ] \Gamma' C^T 
\bar q_1^b (x) \right ) \bar h_v^c (x) , 
\end{equation*}
where~$A$ and~$B$ are two constants satisfying 
the constraint $A + B = 1$ which accounts for 
an arbitrariness in the choice of a local current.  
The variation in $A \in [0,1]$ is adopted 
as a systematic error of numerical estimations.  
Note that the central value $A = B = 1/2$ 
corresponds to the constituent quark model 
picture~\cite{Ball:2008fw}.   
The Dirac matrix~$\Gamma'$ is a suitable structure 
determined by the spin-parity of the baryon, 
in particular, $\Gamma' = \gamma_5$ for baryons 
from the $SU(3)_F$ antitriplet ($j^p = 0^+$) and  
$\Gamma' = \gamma_\|, \, \gamma_\perp$ for 
the $SU(3)_F$-sextet baryons with $j^p = 1^+$.  

In calculations of the correlation functions,  
one tacitly assumes that baryons are bound states 
of quarks which are not free particles inside 
but couple by virtue of the gluonic field.  
So, light quark propagators~$\tilde S_q (x)$, 
being very sensitive to the influence of the 
background gluonic field, should be modified 
accordingly while for the heavy quark this 
effect is sub-dominant and to leading order 
in the heavy-quark mass~$m_Q$ expansion 
can be neglected.   
To take effects of the QCD background inside 
baryons into account, the method of non-local 
condensates~\cite{Mikhailov:1986be,Mikhailov:1991pt,Bakulev:2002hk}
is used. In this approach the light-quark 
propagator can be decomposed into two parts: 
the perturbative~$S_q (x)$ and 
non-perturbative~$\mathcal{C}_q (x)$ ones,  
and the later accumulates an information about the 
background inside the baryon in terms of non-local 
quark condensate~$\langle \bar q (x) q (0) \rangle$.

To obtain the QCD sum rules, it is convenient to make 
the double Fourier transform of the correlation function:  
\begin{eqnarray}
\hspace*{-5mm}
\Pi_{\Gamma\Gamma'} (\omega_1, \omega_2; E) =
i \int_{-\infty}^\infty \frac{d t_1 \, d t_2}{(2 \pi)^2} \, 
e^{i (\omega_1 t_1 + \omega_2 t_2)} \int d^4x \, e^{-iE (v x)} \, 
\langle 0| {\cal O}^\Gamma (t_1, t_2) \, \bar J^{\Gamma'} (x) 
| 0 \rangle 
\nonumber 
\end{eqnarray}
As it is well-known from the QCD-SR analysis 
within the HQET, the heavy-quark condensate 
term is suppressed by~$1/m_Q$ and absent 
in the Heavy-Quark Symmetry limit.  
So, the QCD Sum Rules can be read off after the 
phenomenological and perturbatively calculated   
considerations of the correlation function are 
equated based on the idea of the quark-hadron 
duality~\cite{Shifman:1978bx}:
%
\begin{equation*}
|f_H|^2 \, \psi^\Gamma (\omega, u) \, 
e^{- \bar\Lambda_H/\tau} =
\mathds{B}[\Pi](\omega, u; \tau, s_0) , 
\end{equation*}
%
where symbol~$\mathds{B}$ means the Borel-transform,  
$\bar\Lambda_H = m_H - m_Q$ is the effective baryon 
mass in the HQET, and~$s_0$ is the momentum cutoff 
resulting from applying the quark-hadron duality.  
The explicit QCD-SRs for all the baryonic non-local 
operators can be found in~\cite{Ali:2012pn}.

\begin{figure}[tb] 
\centerline{
\includegraphics[width=0.45\textwidth]{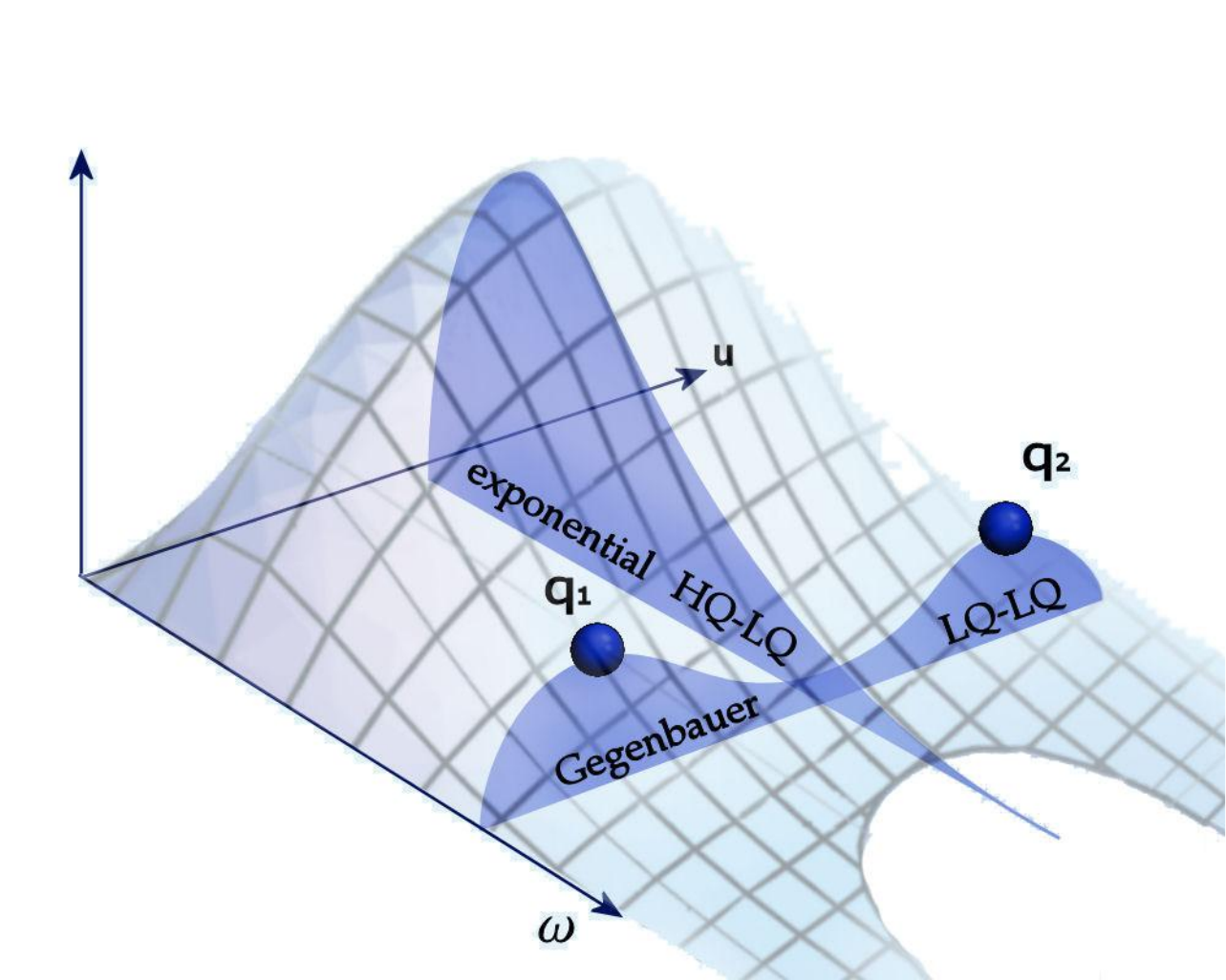}  
} 
\caption{
The general representation of the model functions  
for the heavy-baryon LCDAs with the $\omega$-dependence 
specific for the $B$-meson LCDAs and the $u$-dependence 
in terms of an expansion in the Gegenbauer polynomials 
similar to the ones for the light mesons.     
} 
\label{fig:LCDA-gen-repres}
\end{figure}

The numerical values of first several moments  
of the bottom-baryon LCDAs estimated by the 
QCD-SRs are presented in~\cite{Ali:2012pn}. 
These moments should be matched to the corresponding 
moments of the model functions for the LCDAs. The general 
presentation of the model functions for the $b$-baryon 
LCDAs is governed by their scale evolution and can be 
composed of the exponential part corresponding to the 
heavy-light interaction and the Gegenbauer polynomials 
to the light-light interaction. The order of the polynomials 
is determined by the twist of the diquark system. Motivated 
by the analysis done for the $\Lambda_b$-baryon~\cite{Ball:2008fw},   
the following simple models for the LCDAs  
have proposed~\cite{Ali:2012zza,Ali:2012pn}:     
\begin{eqnarray} 
\tilde \psi_2 (\omega, u) & = & 
\omega^2 u (1 - u) \, \sum_{n = 0}^2  
\frac{a^{(2)}_n}{{\epsilon^{(2)}_n}^4} \, C^{3/2}_n (2 u - 1) \, 
{\rm e}^{- \omega/\epsilon^{(2)}_n} , 
\nonumber \\
\tilde \psi_{3s} (\omega, u) & = & 
\frac{\omega}{2} \, \sum_{n = 0}^2   
\frac{a^{(3)}_n}{{\epsilon^{(3)}_n}^3} \, C^{1/2}_n (2 u - 1) \, 
{\rm e}^{- \omega/\epsilon^{(3)}_n} , 
\nonumber \\
\tilde \psi_{3\sigma} (\omega, u) & = &  
\frac{\omega}{2} \, \sum_{n = 0}^3  
\frac{b^{(3)}_n}{{\eta^{(3)}_n}^3} \, C^{1/2}_n (2 u - 1) \, 
{\rm e}^{- \omega/\eta^{(3)}_n} , 
\nonumber \\
\tilde \psi_4 (\omega, u) & = & 
\sum_{n = 0}^2  
\frac{a^{(4)}_n}{{\epsilon^{(4)}_n}^2} \, C^{1/2}_n (2 u - 1) \, 
{\rm e}^{- \omega/\epsilon^{(4)}_n} .  
\nonumber 
\end{eqnarray}
The qualitative behavior of the twist-2 LCDAs 
is presented in Fig.~\ref{fig:LCDA-gen-repres}. 
The estimates of the parameters entering the 
theoretical models for the heavy-baryon LCDAs 
at the scale $\mu_0 = 1$~GeV can be found 
in~\cite{Ali:2012zza,Ali:2012pn}. 
The $SU(3)_F$-symmetry breaking in LCDAs based 
on taking into account the $s$-quark difference 
from the $u$- and $d$-quarks is estimated 
to be approximately~$15\%$~\cite{Ali:2012zza,Ali:2012pn}.     

\section{Conclusions}
\label{sec:Parkh-conclusions}

A brief introduction into the heavy hadron wave-functions 
is given in the present lecture. The discussion is started 
from the simplest hadronic system~--- the heavy meson which 
is the bound state of the light quark and the heavy antiquark. 
An example of such a system are the $D$- and $B$-mesons 
which are containing the heavy $c$- and $b$-antiquarks, 
respectively. In the HQET, the dynamics in the heavy mesons 
is mainly determined by a motion of the light quark while 
the heavy antiquark can be considered as a static source. 
In many cases, not all the light-quark degrees of freedom 
are equally important and the dominant ones can be separated 
from the sub-dominant. Within this picture, it is possible 
to determine the so-called Light-Cone Distribution Amplitudes 
(LCDAs) of the heavy meson. Analysis shows that there are 
only two LCDAs in the lowest (two-particle) Fock decomposition 
and four more LCDAs in the three-particle heavy-meson state. 
The former two are mainly discussed in this lecture as well 
as their importance in calculations of branching fractions 
of radiative, leptonic and semileptonic $B$-meson decays. 
The last part is devoted to the bottom baryons which are 
the bound states of a heavy quark and two light ones. 
The total sets of the non-local light-ray operators for  
the ground-state heavy baryons with $J^P = 1/2^+$ and 
$J^P = 3/2^+$ are constructed in QCD in the heavy-quark 
limit. Matrix elements of these operators sandwiched 
between the heavy-baryon state and vacuum determine 
the LCDAs of different twist through the diquark current.  
Simple theoretical models for the LCDAs have been proposed 
and are briefly discussed. $SU(3)_F$ breaking effects result 
a correction of order~10\%. Their application, for example,  
to transition matrix elements of heavy baryons is a good 
topic for future theoretical studies.

\section*{Acknowledgements}

I would like to thank the organizers of the Helmholtz 
International Summer School for the invitation, 
skillful organization, and kind hospitality in Dubna.
It is a pleasure to thank Ahmed Ali, Christian Hambrock, 
and Wei Wang for the collaboration on the topic of bottom 
baryons and Anna Kuznetsova for the help with the $B$-meson 
LCDA models. 
I acknowledge the support by the Russian Foundation 
for Basic Research (Project No. 15-02-06033-a).


\begin{footnotesize}

\end{footnotesize}


\end{document}